\documentclass[12pt]{article}
\usepackage{amsmath,amssymb,amsbsy,amsfonts,amsthm,latexsym,amsopn,amstext,amsxtra,euscript,amscd}

\newfont{\teneufm}{eufm10}
\newfont{\seveneufm}{eufm7}
\newfont{\fiveeufm}{eufm5}
\newfam\eufmfam
       \textfont\eufmfam=\teneufm \scriptfont\eufmfam=\seveneufm
       \scriptscriptfont\eufmfam=\fiveeufm
\def\bbbc{{\mathchoice {\setbox0=\hbox{$\displaystyle\rm C$}\hbox{\hbox
to0pt{\kern0.4\wd0\vrule height0.9\ht0\hss}\box0}}
{\setbox0=\hbox{$\textstyle\rm C$}\hbox{\hbox
to0pt{\kern0.4\wd0\vrule height0.9\ht0\hss}\box0}}
{\setbox0=\hbox{$\scriptstyle\rm C$}\hbox{\hbox
to0pt{\kern0.4\wd0\vrule height0.9\ht0\hss}\box0}}
{\setbox0=\hbox{$\scriptscriptstyle\rm C$}\hbox{\hbox
to0pt{\kern0.4\wd0\vrule height0.9\ht0\hss}\box0}}}}
\def\bbbq{{\mathchoice {\setbox0=\hbox{$\displaystyle\rm
Q$}\hbox{\raise
0.15\ht0\hbox to0pt{\kern0.4\wd0\vrule height0.8\ht0\hss}\box0}}
{\setbox0=\hbox{$\textstyle\rm Q$}\hbox{\raise
0.15\ht0\hbox to0pt{\kern0.4\wd0\vrule height0.8\ht0\hss}\box0}}
{\setbox0=\hbox{$\scriptstyle\rm Q$}\hbox{\raise
0.15\ht0\hbox to0pt{\kern0.4\wd0\vrule height0.7\ht0\hss}\box0}}
{\setbox0=\hbox{$\scriptscriptstyle\rm Q$}\hbox{\raise
0.15\ht0\hbox to0pt{\kern0.4\wd0\vrule height0.7\ht0\hss}\box0}}}}
\def\bbbt{{\mathchoice {\setbox0=\hbox{$\displaystyle\rm
T$}\hbox{\hbox to0pt{\kern0.3\wd0\vrule height0.9\ht0\hss}\box0}}
{\setbox0=\hbox{$\textstyle\rm T$}\hbox{\hbox
to0pt{\kern0.3\wd0\vrule height0.9\ht0\hss}\box0}}
{\setbox0=\hbox{$\scriptstyle\rm T$}\hbox{\hbox
to0pt{\kern0.3\wd0\vrule height0.9\ht0\hss}\box0}}
{\setbox0=\hbox{$\scriptscriptstyle\rm T$}\hbox{\hbox
to0pt{\kern0.3\wd0\vrule height0.9\ht0\hss}\box0}}}}
\def\bbbs{{\mathchoice
{\setbox0=\hbox{$\displaystyle     \rm S$}\hbox{\raise0.5\ht0\hbox
to0pt{\kern0.35\wd0\vrule height0.45\ht0\hss}\hbox
to0pt{\kern0.55\wd0\vrule height0.5\ht0\hss}\box0}}
{\setbox0=\hbox{$\textstyle        \rm S$}\hbox{\raise0.5\ht0\hbox
to0pt{\kern0.35\wd0\vrule height0.45\ht0\hss}\hbox
to0pt{\kern0.55\wd0\vrule height0.5\ht0\hss}\box0}}
{\setbox0=\hbox{$\scriptstyle      \rm S$}\hbox{\raise0.5\ht0\hbox
to0pt{\kern0.35\wd0\vrule height0.45\ht0\hss}\raise0.05\ht0\hbox
to0pt{\kern0.5\wd0\vrule height0.45\ht0\hss}\box0}}
{\setbox0=\hbox{$\scriptscriptstyle\rm S$}\hbox{\raise0.5\ht0\hbox
to0pt{\kern0.4\wd0\vrule height0.45\ht0\hss}\raise0.05\ht0\hbox
to0pt{\kern0.55\wd0\vrule height0.45\ht0\hss}\box0}}}}
\def\bbbz{{\mathchoice {\hbox{$\sf\textstyle Z\kern-0.4em Z$}}
{\hbox{$\sf\textstyle Z\kern-0.4em Z$}}
{\hbox{$\sf\scriptstyle Z\kern-0.3em Z$}}
{\hbox{$\sf\scriptscriptstyle Z\kern-0.2em Z$}}}}

\newtheorem{theorem}{Theorem}
\newtheorem{lemma}[theorem]{Lemma}

\def\cA{{\cal A}}

\def\cF{{\cal F}}
\def\cG{{\cal G}}
\def\cH{{\cal H}}

\def\cM{{\cal M}}

\newcommand{\remove}[1]{}
\newcommand{\variety}{\mathcal{V}}
\newcommand{\card}[1]{\left| #1 \right|}

\newcommand{\Beta}{\textrm{B}}

\def \newchi{\widetilde{\chi}}
\def \newOrf{\widetilde{\Orf}}

\def\({\left(}
\def\){\right)}
\def\[{\left[}
\def\]{\right]}
\def\<{\langle}
\def\>{\rangle}
\def\vec#1{\mathbf{#1}}

\def\rf#1{\left\lceil#1\right\rceil}

\def \F{\mathbb{F}}

\def \Z{\mathbb{Z}}

\def \C{\mathbb{C}}

\def\Ors{{\mathcal{O}}}
\def\Orf{{\mathcal{O}_{f}}}

\newcommand{\eqdef}{{\stackrel{\Delta}{=}}}
\newcommand{\abs}[1]{\left| #1 \right|}
\newcommand{\norm}[1]{\left\| #1 \right\|}

\def\e0{\mathbf{\,e}}

\def\cHp{\cG}
\def\cFd{\cF_d}

\newcommand{\ket}[1]{\left| #1 \right \rangle}

\newcommand{\inner}[2]{{\left\langle #1, #2 \right \rangle}}

\begin{document}

\title{\bf{Classical and Quantum Polynomial Reconstruction via
    Legendre Symbol Evaluation}}

\author{
{\sc Alexander Russell} \\
{Department of Computer Science and Engineering}\\
    {University of Connecticut, Storrs, CT 06269 USA} \\
{\tt acr@cse.uconn.edu} \\
\and
{\sc Igor E.~Shparlinski} \\
{Department of Computing}\\
{Macquarie University, Sydney, NSW 2109, Australia} \\
{\tt igor@ics.mq.edu.au}
}

\maketitle

\begin{abstract}
  We consider the problem of recovering a hidden monic polynomial
  $f(X)$ of degree $d \geq 1$ over a finite field $\F_p$ of $p$
  elements given a black box which, for any $x \in \F_p$, evaluates
  the quadratic character of $f(x)$. We design a classical algorithm
  of complexity $O(d^2p^{d + \varepsilon})$ and also show that the
  quantum query complexity of this problem is $O(d)$. Some of our
  results extend those of Wim van Dam, Sean Hallgren and Lawrence Ip
  obtained in the case of a linear polynomial $f(X) = X+ s$ (with
  unknown $s$); some are new even in this case.
\end{abstract}

\section{Introduction}
Let $p\geq 3$ be a prime number and let $\F_p$ denote a finite field
of $p$ elements. We let $\chi$ denote the quadratic character of
$\F_p$, or the {\it Legendre symbol\/} modulo $p$%
; see~\cite{LN}.

Wim van Dam, Sean Hallgren and Lawrence Ip, in the series of
papers~\cite{vDam,vDamHallg,vDamHallgIp,Ip} have considered the
{\it shifted Legendre symbol\/} problem of finding an unknown shift
$s \in \F_p$
given an oracle $\Ors$ which for each $x \in \F_p$ computes $\chi(x +s)$.
They have designed efficient quantum algorithms for the above problem
and its generalisation to characters in residue rings.

The problem is of intrinsic interest and also has strong cryptographic
motivation; sequences of values of quadratic characters have
been considered as sources of cryptographically strong pseudorandom
bits~\cite{AG,BL1,Damg,Ding,HofLin,Maud,MaSa,Pera2,RivSar}.

Here we consider a generalisation of the above problem to polynomials.

For an integer $d$ we let  $\cM_d$ denote the
set of square-free monic polynomials $f(X) \in \F_p[X]$ of degree $d$,
$$
\cM_d \; \eqdef\;  \left\{ f(X) = X^d + s_{d-1} X^{d-1} + \ldots +
s_1 X + s_0 \mid s_i \in \F_p \right\}.
$$
We study the problem of finding $f \in \cM_d$,
given an oracle $\Orf$ which returns  $\chi(f(x))$
for any $x \in \F_p$:
$$
 \Orf: x \mapsto \chi(f(x)).
$$
 It is obvious that the square-freeness condition
is essential because polynomials of the form $f_1(X) = F(X)G_1(X)^2$
and $f_1(X) = F(X)G_2(X)^2$ with $F(X), G_1(X), G_2(X) \in \F_p[X]$
cannot be distinguished by this oracle.

We remark that for the approach of~\cite{vDam,vDamHallg,vDamHallgIp,Ip}
the orthogonality condition
$$
\sum_{x \in \F_p} \chi\((x + a)(x+b)\) =  \left\{ \begin{array}{ll}
p-1, & \quad \mbox{if}\ a = b, \\
-1, & \quad \mbox{if}\ a \neq b,
\end{array} \right.
\qquad a,b \in \F_p,
$$
   appears to be crucial, however, this condition fails for
nonlinear polynomials.
On the other hand, the {\it Weil bound\/}, see~\cite{LN},  provides a certain
approximate analogue of the above identity:
\begin{equation}
\label{eq:PolyCorr}
\sum_{x \in \F_p} \chi\(g(x) h(x)\) =  \left\{ \begin{array}{ll}
p + O(d), & \quad \mbox{if}\ g = h, \\
O(p^{1/2}), & \quad \mbox{if}\ g \neq h,
\end{array} \right. \qquad g,h \in \cM_d.
\end{equation}
Hereafter the implied constants in symbols `$O$'
may depend on $d$ and also, where obvious, on the small positive parameter
$\varepsilon $.

Using this property
we demonstrate that the quantum query complexity of recovering $f$,
given the oracle $\Orf$, is $O(d)$. In contrast, we observe that the
classical query complexity is $\Omega(d \log p)$. Furthermore, we give a
classical algorithm for reconstructing $f$ from $\Orf$, which appears
to be new even in the case of linear polynomials. In fact this
algorithm is also based on the Weil bound.

It is clear that the brute force approach leads to a (classical)
algorithm of complexity $O(p^{d+1 + \varepsilon})$ which is based on
computation and comparison of the $p$-dimensional vectors of the
values of $\chi(f(x))$ and $\chi(g(x))$ for all $x \in \F_p$ and all
$g \in \cM_d$.  A naive use of the Weil bound shows that it is enough
to compute and compare $\chi(f(x))$ and $\chi(g(x))$ only for $1 \leq
x \leq dp^{1/2} \log^2 p$ which leads to an algorithm of complexity
$O(p^{d+1/2 + \varepsilon})$.  We show that using the Weil bound in a
less obvious way one can obtain an $O(p^{d+\varepsilon})$ algorithm.

It  could be relevant to recall the work of Dima Grigoriev~\cite{Grig}
where a somewhat related question is considered for multivariate polynomials
(although the field characteristic is assumed to be small).

It is easy to see that our method applies to multiplicative characters of other
orders and to multivariate polynomials as well.

Finally, it is also easy to see
that we can allow oracles which
return the right value of $\chi(f(x))$ only with some fixed
probability $\gamma >1/2$. We do not pursue this issue in this work, though.

\bigskip

 {\bf Acknowledgement.} \quad
We thank Asma Harcharras for several
useful  discussions.

\section{Preparation}
\label{sec:Preps}

First of all we recall the   Weil  bound
in its classical form given in
Example~12 of Appendix~5 of~\cite{W};
see also Theorem~3 of Chapter~6 in~\cite{Li} and
Theorem~5.41 and comments to Chapter~5 of~\cite{LN}.

\begin{lemma}
\label{le:Weil}
For any $F  \in \cM_d$ which is not a
perfect square of another polynomial,
the bound
$$
\left|\sum_{x \in \F_p} \chi\( F(x)\)\right|\leq  d p^{1/2}
$$
holds.
\end{lemma}

The following  statement is also implied by  the Weil bound
and is  essentially
Theorem~2 of~\cite{MaSa}.

\begin{lemma}
\label{le:WeilShort}
For any integers $M<p$ and any $F \in \cM_d$ which is not a perfect
square of another polynomial, the bound
$$\left|\sum_{x=1}^{M} \chi(F(x))\right| = O\( d p^{1/2} \log p\)$$
holds.
\end{lemma}

We also need a similar statement for multivariate
polynomials.

\begin{lemma}
\label{le:MultWeil}
For any collection of $\ell$ pairwise distinct linear forms
$$
L_\nu\(S_0, \ldots, S_{d-1} \) = S_0 + S_1 c_{1\nu} + \ldots +
S_{d-1} c_{d-1, \nu} + c_{d,\nu},
\qquad \nu = 1, \ldots, \ell,
$$
over $\F_p$
the bound
$$
\left|\sum_{s_0, \ldots, s_{d-1} \in \F_p}
   \chi\(\prod_{\nu=1}^\ell L_\nu(s_0, \ldots, s_{d-1})\)\right|\leq
2 \ell p^{d - 1/2}
$$
holds.
\end{lemma}

\begin{proof} We have
   \begin{eqnarray*}
     \lefteqn{\left|\sum_{s_0, \ldots, s_{d-1} \in \F_p}
         \chi\(\prod_{\nu=1}^\ell L_\nu(s_0, \ldots, s_{d-1})\)\right| }\\
& & \qquad\leq
\sum_{s_1, \ldots, s_{d-1} \in \F_p}\left|\sum_{s_0 \in \F_p}
   \chi\(\prod_{\nu=1}^\ell L_\nu(s_0, \ldots, s_{d-1})\)\right|.
\end{eqnarray*}
Clearly, there are at most
$$
   \frac{(\ell-1)(\ell-2)}{2}p^{d-2} \leq \ell2 p^{d-2}
   $$
   $d-1$-tuples $(s_1\ldots, s_{d-1}) \in \F_p$ for which the values of
$$
s_1 c_{1\nu} + \ldots + s_{d-1} c_{d-1, \nu} + c_{d,\nu},
\qquad \nu = 1, \ldots, \ell,
$$
are pairwise distinct. In this case we estimate the sum over $s_0$ by $p$.
Otherwise we see from Lemma~\ref{le:Weil} that  the sum over $s_0$
does not exceed $\ell p^{1/2}$.
Therefore
$$
\left|\sum_{s_0, \ldots, s_{d-1} \in \F_p}
   \chi\(\prod_{\nu=1}^\ell L_\nu(s_0, \ldots, s_{d-1})\)\right|\leq
\ell p^{d - 1/2} + \ell2 p^{d-1}.
$$
The claimed bound is trivial for $\ell \geq p^{1/2}$, otherwise we have
$\ell p^{d - 1/2} \geq  \ell2 p^{d-1}$ and the result follows.
\end{proof}

We remark that one can also use stronger  bounds based on
the famous results of Pierre Deligne~\cite{Del1,Del2},
however they do improve our final results.

Our next statement gives an upper bound ``on average''
for weighted character sums with polynomials.

\begin{lemma}
\label{le:Average}
For any integers $N \leq p$, $r \geq 1$ and any sequence of real numbers
$\alpha_x$ with $|\alpha_x| \leq 1$, $x =1, \ldots, N$, the bound
$$
\sum_{g \in \cM_d} \left|\sum_{x\in \F_p} \alpha_x \chi\(
g(x)\)\right|^{2r} \leq 4r N^{2r} p^{d-1/2} +
\frac{(2r)!}{r!} N^r p^d
$$
holds.
\end{lemma}

   \begin{proof}   We have
\begin{eqnarray*}
\sum_{g \in \cM_d} \left|\sum_{x\in \F_p}\alpha_x \chi\( g(x)\)\right|^{2r}
& = & \sum_{g \in \cM_d} \sum_{x_1, \ldots, x_{2r} \in \F_p}
\prod_{i=1}^{2r}\alpha_{x_i} \chi\( g(x_i)\)
\\ & = & \sum_{x_1, \ldots, x_{2r} \in \F_p}\prod_{i=1}^{2r}\alpha_{x_i}
\sum_{g \in \cM_d}    \chi\( \prod_{i=1}^{2r} g(x_i)\) \\
& = & \sum_{x_1, \ldots, x_{2r} \in \F_p} \left|
\sum_{g \in \cM_d}    \chi\( \prod_{i=1}^{2r} g(x_i)\) \right|.
\end{eqnarray*}
Assume that $x_1, \ldots, x_{2r} \in [1,N]$ contains $m$ pairs
of equal elements
$$
x_{i_\nu} = x_{j_\nu} , \qquad \nu = 1, \ldots, m,
$$
and $l = 2r-2m$ pairwise distinct elements $y_\nu = x_{k_\nu}$,
$\nu = 1, \ldots, m$.
Then
$$
\sum_{g \in \cM_d}    \chi\( \prod_{i=1}^{2r} g(x_i)\)  =
\sum_{g \in \cM_d}    \chi\( \prod_{\nu=1}^{l} g(y_\nu)\) .
$$
If $l = 0$, which happens for at most
   $$
r! {2r \choose r} N^r =  \frac{(2r)!}{r!} N^r
$$
$2r$-tuples $\(x_1, \ldots, x_{2r}\) \in [1,N]^{2r}$, then the sum over
$g$ is obviously equal to $\card{\cM_d} = p^d$. For $l > 0$ we derive
$$
\sum_{g \in \cM_d}    \chi\( \prod_{\nu=1}^{l} g(y_\nu)\)
= \sum_{s_0, \ldots, s_{d-1} }
\chi\( \prod_{\nu=1}^{l} \(s_0 + s_1y_\nu + \ldots +s_{d-1}
y_\nu^{d-1} + y_\nu^d \)\).
$$
It is easy to verify that because $y_1, \ldots, y_l$ are
pairwise distinct elements of $\F_p$ the linear forms
$$
S_0 +  S_1y_\nu + \ldots +S_{d-1} y_\nu^{d-1} + y_\nu^d , \qquad \nu
= 1, \ldots, m,
$$
satisfy the conditions of Lemma~\ref{le:MultWeil}. Thus for at most $N^{2r}$
remaining $2r$-tuples  $\(x_1, \ldots, x_{2r}\) \in [1,N]^{2r}$
the sum over $g$ is  at most $2l p^{d-1/2} \leq 4r p^{d-1/2}$
\end{proof}

We recall that, using the Horner scheme, for  any $g\in \cM_d$ the value
of $g(x)$ can be computed with $O(d)$ arithmetic operations modulo $p$.
We also recall that polynomial evaluation and computing the quadratic character
can be done in polynomial time  in the standard RAM model of computation.
Explicit and efficient versions of these statements can be found
in~\cite{AHU,vzGG}.

\section{Classical Algorithm}

Here we design an algorithm for the classical
model of computation on a RAM computer. The complexity of our algorithm
can be improved slightly if one uses fast algorithms for finite field
arithmetic and polynomial evaluation, see~\cite{AHU,vzGG}. In particular,
one can replace $p^{\varepsilon}$ by a reasonably small power of $\log p$
and also improve the term $d^2$ in our estimate.

\begin{theorem}
\label{thm:Class} For any fixed $\varepsilon > 0$ and $d\geq 1$, given
an oracle $\Orf$
one can find $f \in \cM_d$ in $O\(d^2 p^{d + \varepsilon}\)$ binary operations.
   \end{theorem}

\begin{proof}
Obviously we can assume that $p$ is
sufficiently large.
Put $M = \rf{d p^{1/2} \log^2 p}$ and $N = \rf{d \log^2p} $.

Using the square-freeness condition
we conclude that for any $g\in \cM_d$ with $g \neq f$ the polynomial
$gf$  is not a prefect square.
Thus  from
Lemma~\ref{le:WeilShort} we see that in this case
$$
\sum_{x =1}^M \chi(g(x) f(x) ) =O(M/\log p)
$$
while for $g=f$ this sum is at least $M-d$.
Using the Horner scheme, for any any $g\in \cM_d$ the value
of $g(x)$ can be computed with $O(d)$ arithmetic operations modulo $p$.
Thus for any polynomial $g \in \cM_d$ the above sum can be evaluated
and the identity $g=f$ can be
verified in $O(d^2p^{1/2 + \varepsilon})$
binary operations. We now show that in fact for all, except
at most $O(p^{d-1/2}\log p)$ polynomials $g \in \cM_d$ one can verify
the identity $g=f$  in  $O(d N \log^2 p)$
binary operations. It is enough to show that
that the inequality
\begin{equation}
\label{eq:LargeCorr}
\left| \sum_{x =1}^N \chi(g(x) f(x) ) \right| \geq N-d
\end{equation}
is possible for at most $O(p^{d-1/2}\log p)$ polynomials $g \in \cM_d$.
Using Lemma~\ref{le:Average} with $\alpha_x = \chi( f(x) )$,
   we see that the number $T$ of polynomials $g \in \cM_d$
with~(\ref{eq:LargeCorr}),
for any integer $r\geq 1$,
satisfies the inequality
$$
T (N-d)^{2r} \leq 4r N^{2r} p^{d-1/2} + \frac{(2r)!}{r!} N^r p^d .
$$
Let $r =\rf{ \log p}$.
We have
$$(N-d)^{2r} \geq N^{2r} (1-d/N)^{2r} \geq N^{2r}/2
$$
for sufficiently large $p$.
Therefore
$$
T \leq 8r  p^{d-1/2} + 2 \frac{(2r)!}{r!} N^{-r} p^d \leq
8r  p^{d-1/2} + 2 (2r)^r  N^{-r} p^d .
$$
Taking into account that $(2r)^r  N^{-r}  = (2r /N)^{-r}  \leq p^{-1/2}$
for our choice of $N$ and $r$, and sufficiently large $p$,
we obtain the desired
statement.
   \end{proof}

\section{Quantum Query Complexity}

As above, we consider the problem of recovering a polynomial $f$ from
an oracle $\Orf: x \mapsto \chi(f(x))$.  An easy counting argument shows that
the classical query complexity is $\Omega(d \log p)$ (it is in fact $\Theta(d
\log p)$): see, for example, van Dam's article~\cite{vDam}, for an
analogous argument. We begin by showing that the quantum query
complexity of this problem is at most $O(d)$. We refer the reader to
accounts by Nielson and Chuang~\cite{NC} and Kitaev~\cite{K} for a
discussion of quantum computation and quantum algorithms.  In
particular, we need the notion of positive operator valued measurement
(POVM) (see, e.g., \cite{Shor:Quantum}, for a discussion which matches
our notation below).

Recall that a POVM $P$ on Hilbert space $\cH$ is a set $\cA$ and a
family $\left\{\vartheta_a \mid a \in \cA\right\}$ of positive
semidefinite operators on $\mathcal H$ with the property that
$$
\sum_{a \in \cA}  \vartheta_a = \iota,
$$
where $\iota$ denotes the identity operator. The result of the
measurement $P$ on the state $\varPsi \in \cH$ is the probability
distribution on $\cA$ where $a \in \cA$ is observed with probability
$\inner{\vartheta_a \varPsi}{\varPsi}$. Note that $\inner{\vartheta_a
\varPsi}{\varPsi} \geq 0$, as $\vartheta_\alpha$
is positive semidefinite, and that
$$
\sum_{a \in \cA}  \inner{\vartheta_a \varPsi}{\varPsi} = \inner{\sum_{a \in
\cA}  \vartheta_a \varPsi}{\varPsi} = \inner{\iota \varPsi}{\varPsi} 
= \norm{\varPsi}^2 = 1.
$$
Note, also, that in the special case when $\vartheta_a = \gamma \pi$ for a
projection $\pi$ and a scalar $\gamma  \in [0,1]$,
$\inner{\vartheta_a \varPsi}{\varPsi} = \gamma
\norm{\pi \varPsi}^2$.

\begin{theorem}
\label{thm:Quant} Let $f$ be a polynomial in $\cM_d$.
If $d \leq p^{1/2-\varepsilon}$ for some fixed $\varepsilon > 0$ then
there exists a quantum algorithm which, after $O(d)$ quantum queries
to $\Orf$, produces a state for which there is a POVM that determines
$f$ with probability at least $1 + O(p^{-1})$.
\end{theorem}

\begin{proof}  Let us put $k = \rf{2(d+1) \varepsilon^{-1}}$. For a prime $p$,
  let $\cHp$ denote a $p$-dimensional Hilbert space with an
  orthonormal basis $\{ \ket{z} \ \mid \ z \in \Z_p\}$.  Initially, by
  applying the Fourier transform to a delta state, we arrive at the
  uniform superposition
  $$
  \varUpsilon \; \eqdef\;  \frac{1}{\sqrt{p}} \sum_{x \in \F_p} 
\ket{x} \in \cHp
  $$
  which is used to query the oracle $O_f$.  Let $\newchi : \F_p \to
  \{ \pm 1\}$ be the function
  $$
  \newchi(x) = \begin{cases}
    \chi(x) & \text{if}\;x \neq 0,\\
    1 & \text{if}\;x = 0.
  \end{cases}
  $$
  We can certainly assume that in fact we are given an oracle
  $\newOrf$ with $\newOrf(x) = \newchi(f(x))$. Then the result of the
  query may be computed into the phases by controlled phase shift
  yielding the state
  $$
  \varPsi_{f} \; \eqdef\; \frac{1}{\sqrt{p}} \sum_{x \in \F_p}
  \newchi(f(x)) \ket{x}.
  $$
  Repeating the process independently $k \geq 1$ times yields the
  tensor product state
  $$
  \varPsi_{f,k}\; \eqdef\; \frac{1}{p^{k/2}} \sum_{\vec{x} \in \F_p^k}
  \left(\prod_{i=1}^k
    \newchi(f(x_i))\right) \ket{\vec{x}} \in \cHp^{\otimes k}
  $$
  where $\vec{x} = (x_1, \ldots, x_k)$,
  $$
  \cHp^{\otimes k} \;\eqdef\; \underbrace{\cHp \otimes \cdots \otimes \cHp}_k,
  $$
  and $\ket{\vec{x}} \;\eqdef\; \ket{x_1} \otimes \cdots \otimes 
\ket{x_k}$.  In
  general, we let $\varPsi_{g,k}$ denote the state that would have arisen
  at this point had we started with the polynomial $g \in \cM_d$.
  Observe that for $g \in \cM_d$, 
$\inner{\varPsi_{g,k}}{\varPsi_{g,k}} = 1$ and,
  furthermore, that for distinct $g, h \in \cM_d$,
  \begin{eqnarray*}
    \abs{\inner{\varPsi_{g,k}}{\varPsi_{h,k}}} &= &\frac{1}{p^k} 
\abs{ \sum_{\vec{x}
        \in \F^k_p} \prod_{i=1}^k \newchi(g(x_i))\newchi(h(x_i))} \\
& =  &\frac{1}{p^k} \prod_{i=1}^k
    \abs{\sum_{z \in \F_p} \newchi(g(z)) \newchi(h(z))}.
  \end{eqnarray*}
  To bound this, we focus on the inner quantity
  $$
  \sigma_{2d} \; \eqdef\; \max_{\substack{g,h \in \cM_{d}\\ g \neq h}}
  \abs{\sum_{z \in \F_p} \newchi(g(z)) \newchi(h(z))}.
  $$
  For a polynomial $g \in \F_p[X]$, let $\variety(g) = \{ x \in \F_p \mid
  g(x) = 0\}$. Recall that $\variety(gh) = \variety(g) \cup \variety(h)$
  and that for a nonzero univariate polynomial $g$,
  $\card{\variety(g)} \leq \deg(g)$.  Considering that $\newchi(f(x)) =
  \chi(f(x))$ for $x \not\in \variety(f)$, we bound $\sigma_{2d}$ as follows:
\begin{eqnarray*}
  \sigma_{2d} &= & \max_{\substack{g,h \in
      \cM_{d}\\ g \neq h}}
  \abs{\sum_{z \in \F_p \backslash \variety(gh)} \chi(g(z))
\chi(h(z)) + \sum_{z \in \variety(gh)} \newchi(g(z))
\newchi(h(z))}\\
  & \leq & \max_{\substack{g,h \in \cM_{d}\\ g \neq h}}
  \left(\abs{\sum_{z \in \F_p  \backslash \variety(gh)} \chi(gh(z))} +
\card{\variety(gh)}\right)\\
  & \leq  & \max_{\substack{g,h \in \cM_{d}\\ g \neq h}} \abs{\sum_{z \in \F_p}
    \chi(gh(z))}  +  2d.
\end{eqnarray*}
Note now that for two distinct elements $g, h \in \cM_d$, the product
$gh$ cannot be a perfect square and from Lemma~\ref{le:Weil} we
conclude that
\begin{equation}
  \label{bound-sigma}
  \sigma_{2d} \leq \max_{g \in \cM_{2d}} \abs{\sum_{z \in \F_p}
    \chi(g(z))} + 2d \leq   d p^{1/2} + 2d \leq  2 d p^{1/2}
\end{equation}
provided that $p > 3$ (otherwise the result is trivial).  Hence for
distinct $g, h \in \cM_d$ we have
\begin{equation}
  \label{almost-ort}
  \abs{\left\langle \varPsi_{g,k},\varPsi_{h,k} \right\rangle} \leq
  \sigma_{2d}^k p^{-k}.
\end{equation}

Now we show that there is a POVM that identifies the polynomial $f$
with probability $1 + O(p^{-1})$. For each $g \in \cM_d$, let
$\pi_{g,k}$ be the projection operator onto the subspace spanned by
$\varPsi_{g,k}$.  As each $\pi_{g,k}$ is a projection operator, it is
positive semidefinite, and we now show that for some $0 < \alpha < 1$ with
$\alpha = 1 + O(p^{-1})$, there is a decomposition of the identity
operator $\iota$ of the form
$$
\iota = \rho + \sum_{g \in \cM_d} \alpha \pi_{g,k}
$$
where $\rho$ and all $\pi_{g,k}$ are positive semidefinite operators on
$\cHp^{\otimes k}$. Note that if $ \varPsi_{f,k}$ is measured according to
this POVM, the ``correct'' index $f,k$ is observed with
probability $\alpha$.

So define $\rho = \iota - \sum_{g \in \cM_d} \alpha \pi_{g,k}$; we
wish to select $\alpha = 1 + O(p^{-1})$ to insure that $\rho$ is
positive semidefinite. It suffices to see that for our choice of
$\alpha$
\begin{equation}
  \label{inequality}
  \norm{ \sum_{g \in \cM_d} \alpha \pi_{g,k}} < 1,
\end{equation}
where $\norm{M}$ denotes
the {\it operator norm\/} of $M$, given by
$$
\norm{M} \; \eqdef\;  \sup_{\varPhi \neq \vec{0}} \frac{\norm{M 
\varPhi}}{\norm{\varPhi}},
$$
this supremum taken over all nonzero vectors $\varPhi$. Note that for a unit
vector $\varPhi \in \cHp^{\otimes k}$,
$$
\sum_{g \in \cM_d} \pi_{g,k} \varPhi = \sum_{g\in \cM_d}
\inner{\varPhi}{ \varPsi_{g,k}}  \varPsi_{g,k}.
$$

Let $\cFd$ be a Hilbert space of dimension $\card{\cM_d}$
with orthonormal basis $\{ \Beta_g \mid g \in \cM_d \}$ and let $\tau :
{\mathcal G}^{\otimes k} \to \cFd$ be the linear operator
$$
\tau\; \eqdef \sum_{g \in \cM_d} \varPsi_g \Beta_g^*;
$$
here $\Beta_g^* : \cFd \to \C$ is the linear functional
$\Beta_g^*: \varPhi \mapsto \langle \varPhi, \Beta_g\rangle$. Then
$$
\tau \tau^{*}(\varPhi ) = \sum_{g,h \in \cM_d} \varPsi_g \Beta_g^*\left(\Beta_h
  \varPsi_h^*(\varPhi)\right) = \sum_{g\in \cM_d} \varPsi_g 
\varPsi_g^*(\varPhi) =  \sum_{g\in \cM_d} \pi_{g,k} \varPhi,
$$
so that $\sum_g \pi_{g,k} = \tau \tau^*$; recalling that $\norm{\tau^*}^2 =
\norm{\tau \tau^*}$, it suffices to suitably upper bound $\norm{\tau^*}$.
So let $\varPhi \in {\mathcal G}^{\otimes k}$ be an element in the span of
$\{ \varPsi_{g,k} \mid g \in \cM_d \}$ and let $\Gamma = \sum_g \gamma_g \Beta_g \in \cF_d$ satisfy
$\tau(\Gamma) = \varPhi$, which is to say that
$$
\varPhi = \sum_{g \in \cM_d} \gamma_g \varPsi_{g,k}.
$$
Observe that 
\begin{equation}
  \label{eqn:coeffs}
  \begin{split}
    \norm{\varPhi}^2 =& \norm{\sum_{g\in \cM_d} \gamma_g \varPsi_{g,k} }^2 =
  \sum_{g,h \in \cM_d} \gamma_g \gamma_h^*
  \inner{ \varPsi_{g,k}}{ \varPsi_{h,k}}  \\
  =& \remove{\sum_{g\in \cM_d} \abs{\phi_g}^2 +
    O\left(\frac{\sigma_{2d}^k}{p^k} \sum_{\substack{ g,h\in \cM_d \\ g \neq
          h}} \abs{\phi_g \phi_h^*}\right) =} \sum_{g\in \cM_d} \abs{\gamma_g}^2
  + O\left(\frac{\sigma_{2d}^k}{p^k} \left( \sum_{g\in \cM_d}
      |\gamma_g|\right)^2\right)    \\
  =& \norm{\Gamma}^2 + O\left(\frac{\sigma_{2d}^k}{p^k} \( p^d \norm{\Gamma}^2 \)\right)
  = \( 1 + O\left(p^{d-k} \sigma_{2d}^k\right) \) \norm{\Gamma}^2 ,
\end{split}
\end{equation}
by the Cauchy--Schwarz inequality. With $\varPhi$ expressed in this
way, we expand $\norm{\tau^*\varPhi}$ as follows:
\begin{equation}
  \label{eqn:coeff-bound}
  \begin{split}
    \norm{\tau^* \varPhi}^2 =& \sum_{g \in \cM_d} \abs{\inner{\varPhi}{\varPsi_{g,k}}}^2 = \sum_{g \in \cM_d}
    \abs{\inner{\sum_{h \in \cM_d} \gamma_h \varPsi_{h,k} }{\varPsi_{g,k}}}^2\\
    =& \sum_{g \in \cM_d}
    \abs{\gamma_g + \sum_{h \neq g} \gamma_h \inner{\varPsi_{h,k}
      }{\varPsi_{g,k}}}^2.
  \end{split}
\end{equation}
Recalling the inner product bounds of~(\ref{almost-ort}), for
any $g \in \cM_d$ we must have
\begin{equation}\label{eqn:error-terms}
\abs{\sum_{\substack{h \in \cM_d\\ g \neq h}} \gamma_h
\inner{\varPsi_{h,k}}{\varPsi_{g,k}}} \leq \sigma_{2d}^k p^{-k} \sum_{h \in \cM_d} \abs{\gamma_h} \leq \sigma_{2d}^k p^{-k} \sqrt{p^d}
\norm{\Gamma},
\end{equation}
again by the Cauchy--Schwarz inequality. Finally, considering that
$\norm{\alpha + \beta} \leq \norm{\alpha} + \norm{\beta}$, we conclude
from~(\ref{eqn:coeff-bound}) and~(\ref{eqn:error-terms}) that
\begin{align*}
  \norm{\tau^* \varPhi} \leq \norm{\Gamma} + p^{d-k}\sigma_{2d}^k
  \norm{\Gamma} = \left(1 + p^{d-k}\sigma_{2d}^k\right)
  \norm{\Gamma}
\end{align*}
and, from (\ref{eqn:coeffs}), that
$$
\norm{\tau^* \varPhi} \leq \left(1 + O\left(p^{d-k}\sigma_{2d}^k\right)\right)
\norm{\varPhi}.
$$
Hence
$$
\norm{ \sum_{g\in \cM_d} \pi_{g,k}} \leq 1 + O\left(p^{d-k}\sigma^k_{2d}\right).
$$
We can assume that $p > 2^{2/ \varepsilon}$ and hence that $2
\leq p^{\varepsilon/2}$, because otherwise the result is trivial. Then
by~(\ref{bound-sigma}) we have
$$
p^{d-k} \sigma_{2d}^k \leq p^{d-k} \(2 d p^{1/2} \)^k \leq
p^{d-k} p^{(1 - \varepsilon/2)k} = p^{d - k \varepsilon/2} \leq p^{-1},
$$
because of our choice of $k$.
We obtain
$$
\norm{\sum_{g\in \cM_d}\pi_{g,k}} \leq 1 + O\(p^{-1}\),
$$
and are guaranteed that~(\ref{inequality}) holds (provided that $p$
is large enough) for some $\alpha= 1 + O(p^{-1})$ (recall that $(1 +
\delta)^{-1} = 1 + O(\delta)$).  Thus the above POVM determines $f$
with probability $\alpha = 1 + O(p^{-1})$.
\end{proof}

\end{document}